\documentclass[aps,prl,twocolumn,showpacs,superscriptaddress,groupedaddress,nofootinbib,floatfix]{revtex4-1}

\usepackage{graphicx,psfrag}
\usepackage{mathrsfs}
\usepackage{amsmath,amsfonts,amssymb}
\usepackage{multirow}
\usepackage{comment}
\usepackage{ulem}
\usepackage{multirow}
\usepackage{hyperref}

\newcommand{\be}{\begin{equation}}
\newcommand{\ee}{\end{equation}}
\newcommand{\bea}{\begin{eqnarray}}
\newcommand{\eea}{\end{eqnarray}}
\newcommand{\bel}{\begin{align}}
\newcommand{\eel}{\end{align}}

\def\Msun{{\rm M_{\odot}}}
\def\Mmax{{M_{\rm max}^{\rm TOV}}}

\def\Rmax{{R_{\rm max}^{\rm TOV}}}
\def\rhomax{{\rho_{\rm max}^{\rm TOV}}}
\def\RK{{R_{f_2}}}
\def\rhosat{{\rho_{\rm sat}}}
\def\phipm{{\phi_{\rm PM}}}

\def\GMc2{{\rm G M_{\odot} c^{-2}}}

\def\bajes{{\scshape bajes}}
\def\gwbench{{\scshape gwbench}}
\def\nrpm{{\scshape nrpm}}

\def\kt2{{\kappa^{\rm T}_2}}
\def\ie{{\it i.e.}}
\def\eg{{\it e.g.}}

\usepackage{pifont} 

\usepackage{color}
\definecolor{cyan}{rgb}{0,0.9,0.9}
\definecolor{orange}{rgb}{0.9,0.5,0}
\definecolor{magenta}{rgb}{1,0,1}
\definecolor{purple}{rgb}{0.8,0.4,0.8}
\definecolor{gray}{rgb}{0.8242,0.8242,0.8242}

\begin{document}

\title{Constraints on the maximum densities of neutron stars\\
	 from postmerger gravitational waves with third-generation observations}

\author{Matteo \surname{Breschi}$^{1}$}
\author{Sebastiano \surname{Bernuzzi}$^{1}$}
\author{Daniel \surname{Godzieba}$^{2}$}
\author{Albino \surname{Perego}$^{3,4}$}
\author{David \surname{Radice}$^{5,2,6}$}

\affiliation{${}^{1}$Theoretisch-Physikalisches Institut, Friedrich-Schiller-Universit{\"a}t Jena, 07743, Jena, Germany}
\affiliation{${}^{2}$Department of Physics, The Pennsylvania State University, University Park, PA 16802, USA}
\affiliation{${}^{3}$Dipartimento di Fisica, Universit\`{a} di Trento, Via Sommarive 14, 38123 Trento, Italy}
\affiliation{${}^{4}$INFN-TIFPA, Trento Institute for Fundamental Physics and Applications, via Sommarive 14, I-38123 Trento, Italy}
\affiliation{${}^{5}$Institute for Gravitation \& the Cosmos, The Pennsylvania State University, University Park, PA 16802, USA}
\affiliation{${}^{6}$Department of Astronomy \& Astrophysics, The Pennsylvania State University, University Park, PA 16802, USA}

\date{\today}

\begin{abstract}
  Using data from 289 numerical relativity simulations of 
  binary neutron star mergers, we identify, for the first time, a robust
  quasi-universal relation connecting the postmerger peak
  gravitational-wave frequency and the value of the density at the
  center of the maximum mass nonrotating neutron star.
  This relation offers a new possibility for precision
  equation-of-state constraints with next-generation ground-based 
  gravitational-wave interferometers.
  Mock Einstein Telescope observations of fiducial events indicate
  that Bayesian inferences 
  can constrain the maximum density to ${\sim}15\%$ ($90\%$
  credibility level) for a single signal at the minimum
  sensitivity threshold for a detection. 
  If the postmerger signal is included in a full-spectrum
  (inspiral-merger-postmerger) analysis of such signal,
  the pressure-density function can be tightly constrained up to the
  maximum density, 
  and the maximum neutron star mass can be
  measured with an accuracy better than $12\%$ ($90\%$ credibility level).

\end{abstract}


\maketitle

\paragraph{Introduction.---}
Postmerger (PM) gravitational-waves (GWs) emitted from binary
neutron star (BNS) merger remnants are unique probes for the stars'
nuclear equation of state (EOS) at extreme-densities.
During merger, the neutron stars (NSs) fuse together
to form a remnant NS with maximum densities up to two times the
component's mass reaching ${\sim} 3{-}6 \rhosat$ where $\rhosat
\approx 2.7{\times}\,10^{14}\,{\rm g \, cm}^{-3}$ is the nuclear
saturation density. If the remnant does not promptly collapse to 
black hole, numerical relativity (NR) simulations predict a loud
GW transient emitted on dynamical timescales of tens
of milliseconds with a complex signal morphology and a characteristic 
peak frequency $f_2\sim2{-}4~{\rm kHz}$, \eg~\cite{Shibata:2006nm,Hotokezaka:2013iia,Bauswein:2015yca,
	Takami:2014zpa,Bernuzzi:2015rla,Lehner:2016lxy}.
These kiloHertz GW transients have not been detected in the two BNS
events GW170817 and GW190425 due to the insufficient sensitivity of the GW detectors at
those frequencies~\cite{TheLIGOScientific:2017qsa,Abbott:2017dke,Abbott:2018wiz,Abbott:2018hgk,Abbott:2020uma}.
A first detection is however possible (and expected) with
third-generation~\cite{Hild:2010id,Punturo:2010zza} and others
proposed detectors that specifically target the kiloHertz frequencies~\cite{Ackley:2020atn}.
In view of this observational scenario, it is of central importance
to determine what kind of information can be extracted from PM signals
and to which accuracy.

The signal-to-noise (SNR) detection threshold for a PM signal has been
recently studied using different approaches
\eg~\cite{Clark:2015zxa,Chatziioannou:2017ixj,Torres-Rivas:2018svp,
	Tsang:2019esi,Breschi:2019srl,Easter:2018pqy,Easter:2020ifj}. These studies found that SNRs ${\gtrsim}7-8$ at kiloHertz frequencies are
typically necessary to confidently claim the detection of a BNS PM signal. These PM SNRs correspond to 
loud inspiral-merger signals with SNR ${\gtrsim}150$ that, in turn, 
would provide rather accurate measurements of the binary mass and tidal
polarizability parameters, \eg~\cite{Damour:2012yf,Gamba:2020wgg,Chatziioannou:2021tdi}.
While the existence of an underlining, unique EOS for the NS matter
allows to estimate the EOS at {\it all} densities from GW observations, 
inspiral-merger frequencies are mostly
informative at the largest densities of the binary components
\cite{Abbott:2018exr,Landry:2018prl, Agathos:2019sah}.
In contrast, PM frequencies are expected to be mostly informative of
the extreme-densities reached by the remnant,
although BNS PM modeling for GW inference is non-trivial.
The main approach is to employ
EOS-insensitive relations connecting the peak frequency $f_2$,
which is the most robust feature of the PM spectrum as predicted by
simulations, and equilibrium properties of NSs like the radius at fiducial
masses \cite{Bauswein:2012ya,Bauswein:2015vxa}, averaged binary
compactnesses \cite{Takami:2014tva}, or the binary's tidal coupling
constant $\kt2$ \cite{Bernuzzi:2015rla}.
For example, $\kt2$ can be measured from $f_2$ to
within a factor of two (${\sim}20\%$) at PM SNR ${\sim}8$ (15)
\cite{Easter:2020ifj}, but the inspiral signal of the same event would deliver a
measurement orders of magnitude more accurate
\cite{Breschi:2019srl}, thus providing more stringent constraints on
the pressure-density EOS function. 
A PM detection can typically provide only bounds on the maximum NS
mass $\Mmax$ \cite{Bauswein:2013jpa,Agathos:2019sah} 
[solution of the Tolman-Oppenheimer-Volkoff (TOV) equations];
although,
multiple PM detections could provide a few percent measurement
\cite{Bauswein:2014qla} complementary to that from an inspiral EOS
inference \cite{Abbott:2018exr,Landry:2018prl,Essick:2019ldf}. 
The minimum NS radius $\Rmax=R(\rhomax)$, 
\ie~the radius of a NS at maximum density $\rhomax$,
is a key information to constrain the NS mass-radius
diagram. It can be extracted from a quasiuniversal relation with the peak PM frequency
\cite{Bauswein:2014qla} and measured with an uncertainty of
${\sim}4\,{\rm km}$ at PM SNR 12 \cite{Breschi:2019srl,Breschi:2021wzr}.
Stronger EOS constraints from BNS mergers will be possible by
combining GW observations with other messengers,
\eg~\cite{Margalit:2017dij,Radice:2017lry,Bauswein:2017vtn,Coughlin:2018miv,Raaijmakers:2019dks,Dietrich:2020efo,Essick:2020flb,Breschi:2021tbm}.

In this {\it Letter}, we propose a new quasiuniversal relation
connecting the NS maximum density to the peak PM frequency. 
The existence of a EOS-insensitive $\rhomax(f_2)$ relation was 
suggested by previous work \cite{Bauswein:2013jpa,Bauswein:2014qla,Lioutas:2021jbl}, 
but the relation obtained there involves 
the binary configurations with the largest possible mass 
that does not promptly collapse to black hole. 
Here, by introducing an effective Keplerian radius associated to $f_2$, we propose a more general approach that can be applied to any binary
and delivers a strong constrain already with a single detection.
Employing full Bayesian PM analyses of mock signals and incorporating the
expected information from the inspiral-merger signal,
we demonstrate that a single GW BNS detection with the Einstein Telescope at
the PM detection threshold can already deliver a ${\sim}15$\% measurement of
$\rhomax$ and $\Mmax$. 
The novel method shows
the capabilities of next-generation detectors 
and introduce the possibility for direct and accurate constraints 
on the high-density EOS properties
from observational GW data of BNS mergers.

\begin{figure}[t]
  \centering 
  \includegraphics[width=0.49\textwidth]{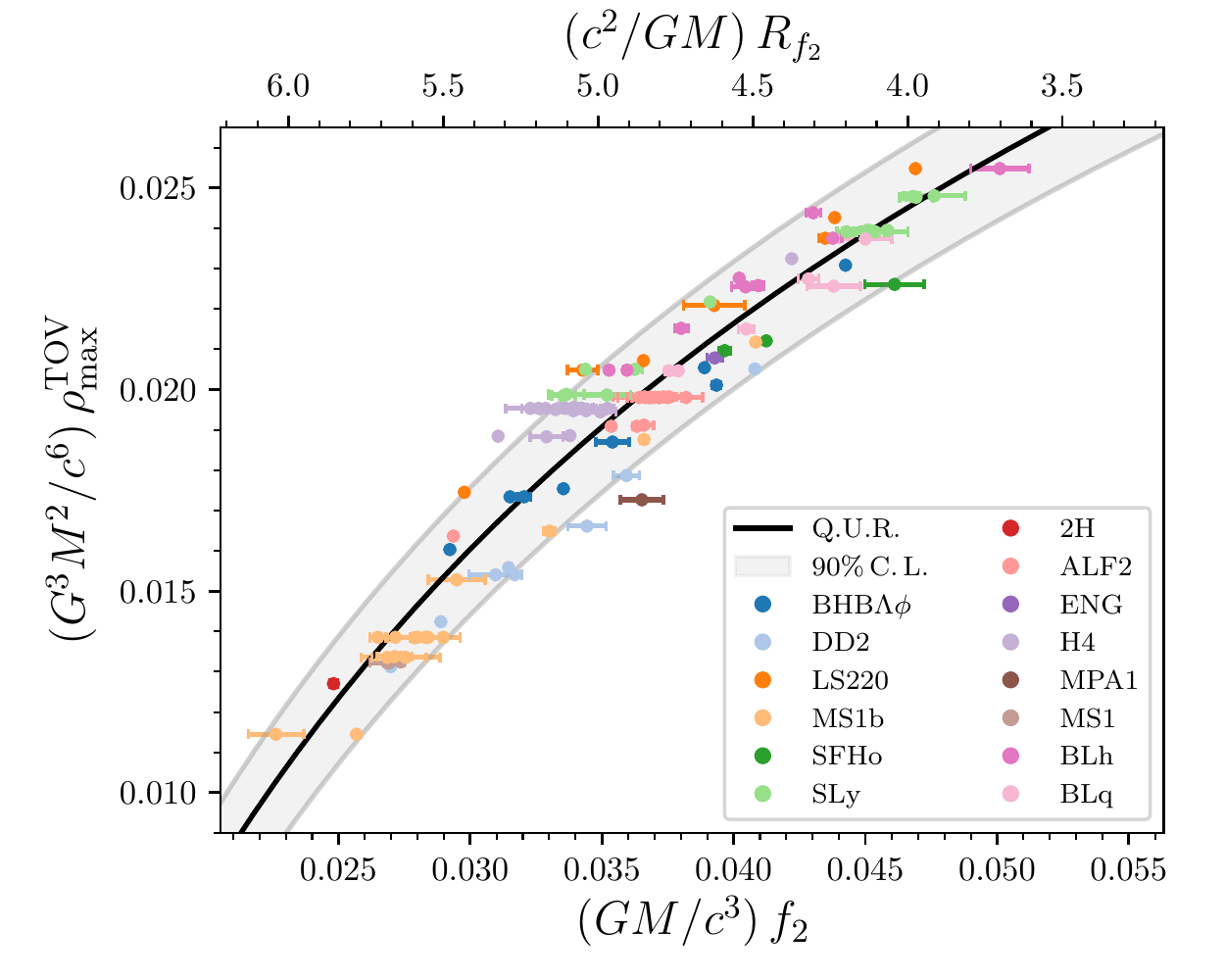}
  \caption{Empirical relation (black line) for 
    the maximum central density $\rhomax$
    of a non-rotating NS as function of
    the Keplerian radius $\RK$
    (top axis) 
    	and the PM peak frequency $f_2$ (bottom axis).
    The colored markers show the data extracted from 
    289 NR simulations with 14 EOSs. 
    Each marker corresponds to a different binary configuration
    and the error bars are computed using different numerical resolutions
    (when available).
    The shadowed area reports the 90\% credibility
    region of the fit. }
  \label{fig:maxfit}
\end{figure}

\paragraph{Effective Keplerian radius.---}
The PM peak frequency $f_2$ can be associated to an effective remnant
radius using Kepler's law, 
\be
\label{eq:rk}
\RK^3 = \frac{G M}{(\pi f_2)^2 }\,,
\ee
where $G$ is the gravitational constant and
$M=m_1+m_2$ is the total mass of the binary.
While this quantity has no direct physical interpretation in terms of
the remnant properties, it correlates to the maximum central density
$\rhomax$ of a non-rotating equilibrium NS with a weak dependence on
the EOS. Figure~\ref{fig:maxfit} shows the approximate
quasiuniversal relation $\rhomax(\RK)$ for a NR sample of 289
simulations of the CoRe
collaboration~\cite{Dietrich:2018phi,Radice:2018pdn,
	Bernuzzi:2020txg,Nedora:2020pak,Prakash:2021wpz}. To explore EOS variation, the 
simulations were performed with 14 different EOSs including 
piecewise polytropes~\cite{Engvik:1995gn,Mueller:1996pm,Alford:2004pf,Lackey:2005tk,Read:2008iy}, 
5 finite-temperature nucleonic models~\cite{Lattimer:1991nc,Douchin:2001sv,Typel:2009sy,Steiner:2012xt,Logoteta:2020yxf}, one
finite-temperature hybrid model accounting for deconfined quark
matter~\cite{Logoteta:2020yxf,Prakash:2021wpz}, and one
finite-temperature hadronic model with $\Lambda$ hyperons~\cite{Banik:2014qja,Radice:2016rys}.
The majority of the sample (${\sim}85\%$) is simulated using microphysics, neutrino transport and a subgrid model for magnetohydrodynamics (MHD) turbulence, that are sufficient to fully capture the features of the PM signal modeled here. For example, the highest-to-date resolution simulations of \cite{Kiuchi:2017zzg} suggest that the MHD effective viscosity is small and does not affect the GWs; our subgrid model is tuned to those simulations. Neutrino induced bulk viscosity might damp radial oscillations of the remnant (which cause secondary peaks in the spectrum) and will not impact $f_2$. Moreover, recent simulations with improved neutrino transport suggest that neutrino bulk viscosity is too small to have a significant impact on dynamics and GW emission \cite{Radice:2021jtw}.

The sample includes binaries with NS masses in the range
$1{-}2\,\Msun$ and dimensionless spin magnitudes $\leq 0.2$.
The relation $\rhomax(\RK)$ fits to the expression:
\be
\label{eq:rhofit}
 \rhomax =  \frac{ a_0 \, c^6}{G^3M^2} \left[1 + a_1 \left(\frac{c^2 \RK}{G M} \right)^{1/4}\right]\,,
\ee 
where $c$ is the speed of light and $(a_0,a_1)=(0.135905,-0.59506)$
are determined by a standard least-squared minimization method at
$\chi^2 = 0.016$. 
The standard deviations of the calibrated coefficients $a_0$
				and $a_1$ are respectively equal to $1.35\%$ and $0.22\%$,
				corresponding to a relative error of $5.8\%$ on the final prediction.
		
The r.h.s. of Eq.~\eqref{eq:rhofit} can be entirely determined from
the measurement of the binary mass $M$ and the PM peak 
frequency $f_2$. 
Writing the Keplerian radius $\RK$ 
				in terms of $f_2$,
				one can get a direct relation between the maximum
				density $\rhomax$ and the PM frequency, as
			  \be
			\rhomax =  \frac{a_0c^6}{G^3M^2}\left[ 1+a_1\left( \frac{c^3}{\pi G M f_2}\right)^{1/6} \right]\,. 
			\ee
Hence, the maximum density $\rhomax$ can be
best inferred from a full-spectrum BNS GW observation. We next
discuss the potential accuracy of such measurement using a mock Bayesian
inference study based on match filtering techniques and a
NR-informed analytical model for the PM signal.

\paragraph{Mock inference study.---}
We consider the Einstein Telescope (ET) 
design~\cite{Hild:2010id,Punturo:2010zz} and assume the detector is
composed by three interferometers with triangular shape and power spectral
density (PSD) configuration D~\cite{Hild:2010id}.
Mock signals are simulated from two fiducial equal-mass non-spinning
BNS: a binary with mass $M=2.73\,\Msun$ and (a ``stiff'') EOS DD2~\cite{Typel:2009sy} 
and a binary with mass $M=2.6\,\Msun$ and (a ``soft'') EOS SLy~\cite{Douchin:2001sv}.
Merger and PM waveforms for these BNSs from NR were already discussed 
in Refs.~\cite{Perego:2019adq,Breschi:2019srl} and are
employed for the PM analyses. The fiducial binaries are placed at different
distances between $80$
and $200~{\rm Mpc}$; the injected signals have
PM SNRs ranging from ${\sim}6$ to $14$, corresponding to
total SNRs from $110$ to ${\sim}260$.
For our case study, we perform full Bayesian analyses of the PM signals
at different SNR and employ the Fisher matrix approach to estimate the
uncertainties of the parameters measured from the inspiral signals.

Bayesian PM analyses are performed using the {\bajes}
pipeline~\cite{Breschi:2021wzr} and an updated version of the
{\nrpm} model for BNS PM signals~\cite{Breschi:2019srl}.
NR data are injected in a segment of $1~{\rm s}$ with a sampling rate
of $16~{\rm kHz}$. The sampling is 
performed with {\scshape dynesty}~\cite{Speagle:2020} with 3200 live
points analyzing the frequency region from $1~{\rm kHz}$ to $8~{\rm
  kHz}$.
{\nrpm} 
is calibrated on a subset of the numerical data shown in Fig.~\ref{fig:maxfit} (see \cite{Breschi:2019srl}) 
and it
employs a simple analytical
prescription for the remnant emission to capture the three peak
frequencies and damping times of typical PM signals. 
These (complex)
frequencies, here collectively indicated as $Q^{\rm NR}$, are fully
determined by the intrinsic binary properties using NR-informed quasiuniversal relations in
terms of the tidal coupling constant $\kt2$ and the symmetric mass
ratio $\nu=m_1m_2/M^2$~\cite{Bernuzzi:2015rla,Zappa:2017xba,Breschi:2019srl}.
The prior distribution is assumed to be uniform in the 
mass components, 
spanning the ranges $M\in [1.5,6]~\Msun$ and $m_1/m_2\in [1,2]$, 
and in the individual NS quadrupolar tidal
polarizability parameters $\Lambda_i\in [0,5000]$ for $i=1,2$. 
Spin magnitudes are kept fixed to
zero. The extrinsic parameters are treated as discussed in
\cite{Breschi:2021wzr}, with volumetric prior for the luminosity
distance in the range $[5,500]~{\rm Mpc}$. 
The likelihood function is analytically marginalized over reference
time and phase. 

In this work, {\nrpm} is extended with two sets of additional
parameters that are determined by the Bayesian inference.
A first set of three parameters $(\alpha,\beta, \phipm)$ is introduced
to enhance the flexibility of the model and to improve fitting factors
to the PM signals with complex morphology.
$\alpha$ is a damping time for the PM
bursts (see Eq.~(9c) of \cite{Breschi:2019srl});
$\beta$ accounts for a linear contribution in the frequency evolution
as discussed in \cite{Easter:2020ifj};
and $\phipm$ is an additional phase-shift corresponding to the first
amplitude minimum \cite{Breschi:2019srl}. 
The prior distribution of these parameters is chosen uniformly in $\alpha^{-1}\in [1,1000]$,
$\beta \in [10^{-4}, -10^{-4}]$ and $\phipm\in[0,2\pi]$.
A second set of recalibration parameters $\delta_Q$ is
introduced to account for the uncertainties of each quasiuniversal
relation $Q^{\rm NR}(\kt2,\nu)$. Specifically, we map $Q^{\rm NR}\mapsto Q^{\rm NR}(1+\delta_Q)$
and treat $\delta_Q$ similarly to standard calibration parameters in GW
analyses, \eg~\cite{Vitale:2011wu}. The prior distribution for
the  $\delta_Q$ is assumed to be normal with zero mean and variance prescribed by the
residuals of the quasiuniversal relations.

This approach aims at combining the strenghts of the {\nrpm} templated analyses, built on quasiuniversal relations, with those of more agnostic analyses using minimal assumptions about the signal morphology, \eg~\cite{Chatziioannou:2017ixj,Easter:2020ifj}. In particular, the improved {\nrpm} model used here improves the fitting factors by an order of magnitude with respect to \cite{Breschi:2019srl}, and can detect PM GW signals at a SNR comparable to unmodeled approaches~\cite{Tsang:2019esi,Easter:2020ifj}. While relying on a larger number of parameters than unmodeled methods, {\nrpm} can deliver a measure of the intrisic parameters independently on pre-merger analyses, and produces narrower $f_2$ posteriors for the same PM SNR. The inference of the tidal coupling constant is also improved: comparing to \cite{Easter:2020ifj}, $\kt2$ is inferred to ${\sim}30\%$ for PM SNR ${\sim}12$ instead of ${\sim15}$. We also note that unmodeled analyses rely on analogous quasiuniversal relations to infer $\kt2$, hence should also consider recalibration parameters.

Fisher matrix analyses of the inspiral-merger signals are performed with
{\gwbench}~\cite{Borhanian:2020ypi}. Considering the large SNR of these
signals, the Fisher matrix provides an approximation of
the posterior distributions of the binary mass and tidal parameters
that is sufficiently accurate for our purposes. For instance, in the
lowest SNR case of $110$, the uncertainty on the binary mass measured
from the inspiral signal is ${\sim}10^{-4}\,\Msun$ and the uncertainty on the
quadrupolar tidal polarizability parameters is ${\sim}2\%$. 
These results imply,
as expected, that the inference on $\rhomax$ via Eq.~\eqref{eq:rhofit} has
uncertainties dominated by the measurement of $f_2$ from the PM signal
and by recalibration errors.

\begin{figure}[t]
	\centering 
	\includegraphics[width=0.49\textwidth]{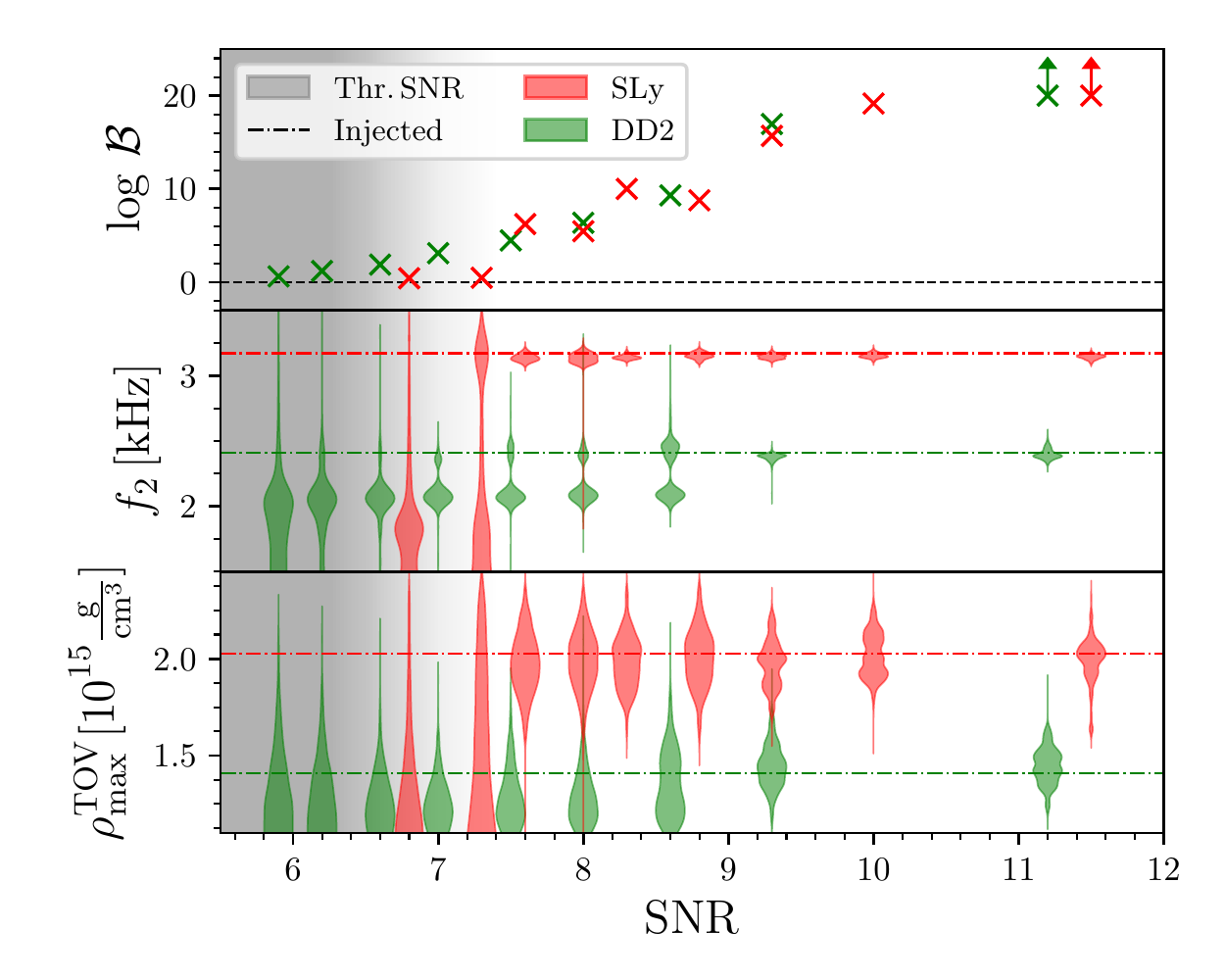}
	\caption{Bayes' factors and posterior distributions
					measured in the mock inference study
					as functions of the PM SNR.
					The DD2 case is reported in
                                        green, 
					the SLy case is reported in red.
					The top panel shows the recovered Bayes' factors $\log\mathcal{B}$.
				 	Central and bottom panels report the 
			 		posterior distributions respectively of $f_2$ and $\rhomax$
		 			and the dashed-dotted lines denote the injected values. 
 					A gray shaded region shows the
                                        detectability
                                        threshold.
     }
	\label{fig:rhomax_snr}
\end{figure}

\begin{figure}[t]
  \centering 
  \includegraphics[width=0.49\textwidth]{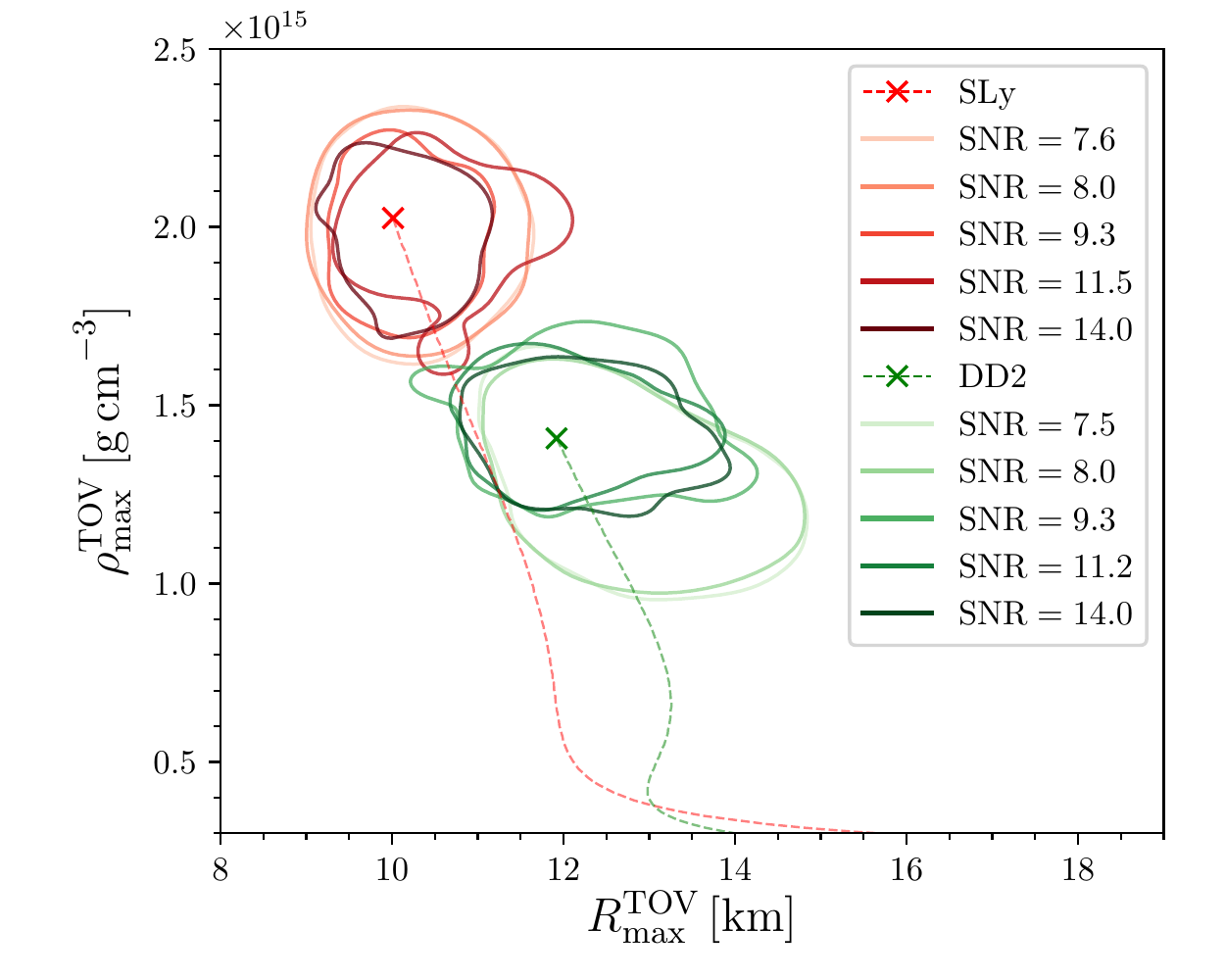}
  \caption{Posterior distributions for
    $\left(\rhomax, \Rmax\right)$
    compared to the EOS of the injected 
    signals.
    The contours show the $90\%$
    credibility regions.
    Green and red lines correspond
    respectively to the results obtained with
    DD2 and SLy EOSs and  
    color intensities 
    refer to the injected SNRs of the PM data.
    The dashed lines show the 
    central density as function
    of the NS radius and the 
    crosses denote the maximum-density values.}
  \label{fig:eos_rhomax}
\end{figure}

\paragraph{Maximum density constraint.---}
Our analysis indicates that PM signals can be detected with SNRs as
low as PM SNR threshold of ${\sim}7.5$ (total SNR ${\sim}140$). As
shown in the top panel of 
Fig.~\ref{fig:rhomax_snr}, the SNR threshold corresponds to a
Bayes' factor of $\log\mathcal{B} \gtrsim 5$ in favor of the signal
(vs. noise) hypothesis.
This detection threshold is comparable to those
obtained with minimally modeled PM
waveform, \eg~\cite{Easter:2020ifj}, and improves 
over the previous {\nrpm} results \cite{Breschi:2019srl} as a
consequence to the enhanced flexibility of our fully NR-informed model.
Figure~\ref{fig:rhomax_snr} (middle panel) also shows the posterior of
$f_2$ at different SNRs. The $90\%$ credibility interval of the $f_2$
measurement is of order of ${\sim}12\%$ at the detectability threshold and
decreases to ${\sim}3\%$ at the largest PM SNR ${\sim}12$.
The DD2 posteriors show bimodalities for SNR ${\lesssim}9$ due
to a loud first peak in the spectrum (the subdominant frequency peak
at frequency $f<f_2$), but this systematic vanishes for increasing SNR.

The posterior distribution for $\rhomax$ is obtained by combining the
posteriors of $M$ from the inspiral analysis and the posteriors of
$f_2$ from the PM analysis.
The uncertainties in the fit of Eq.~\eqref{eq:rhofit} are taken into account
with a resampling similar to the one used for the recalibration
parameters (see also \cite{Breschi:2021tbm}).
Analogously to the $f_2$ measurement, the true
value is recovered at the detectability threshold with an error of
${\sim}15\%$ at the $90\%$ credibility level, as shown in the bottom
panel of Fig.~\ref{fig:rhomax_snr}.
Notably, the inference on $\rhomax$ hits the theoretical uncertainties
on the quasiuniveral relation in Eq.~\eqref{eq:rhofit} at SNR
${\sim}10$.
A more precise measure is either not possible, 
because at that level the relation $\rhomax(\RK)$ becomes
EOS-dependent, or it requires a more precise quasiuniversal relation
from improved simulations.

In order to further illustrate the accuracy of this approach in
constraining high-density NS properties, we consider the
simultaneous measure of $\rhomax$ and of the minimum NS radius $\Rmax$. Here, the
posteriors of $(M,f_2)$ are mapped into $\left(\rhomax, \Rmax\right)$
using Eq.~\eqref{eq:rhofit} and the quasiuniversal relation for
$\Rmax$ in Eq.~(23) of \cite{Breschi:2019srl}.
As above, the uncertainties of the quasiuniversal relation are taken
into account by sampling on appropriate recalibration parameters.
The result is shown in Fig.~\ref{fig:eos_rhomax}: 
the injected values, denoted with crosses, confidently lay
	 within the 90\% credibility regions of the recovered posterior distributions. 
At the detection threshold, the uncertainty on the 
radius $\Rmax$ are of the order of $30\%$ ($90\%$ credibility level).
Such precision is sufficient to distinguish soft and stiff EOSs within the 90\% credibility regions, as 
illustrated using the SLy and DD2 EOSs.


\begin{figure}[t]
	\centering 
	\includegraphics[width=0.49\textwidth]{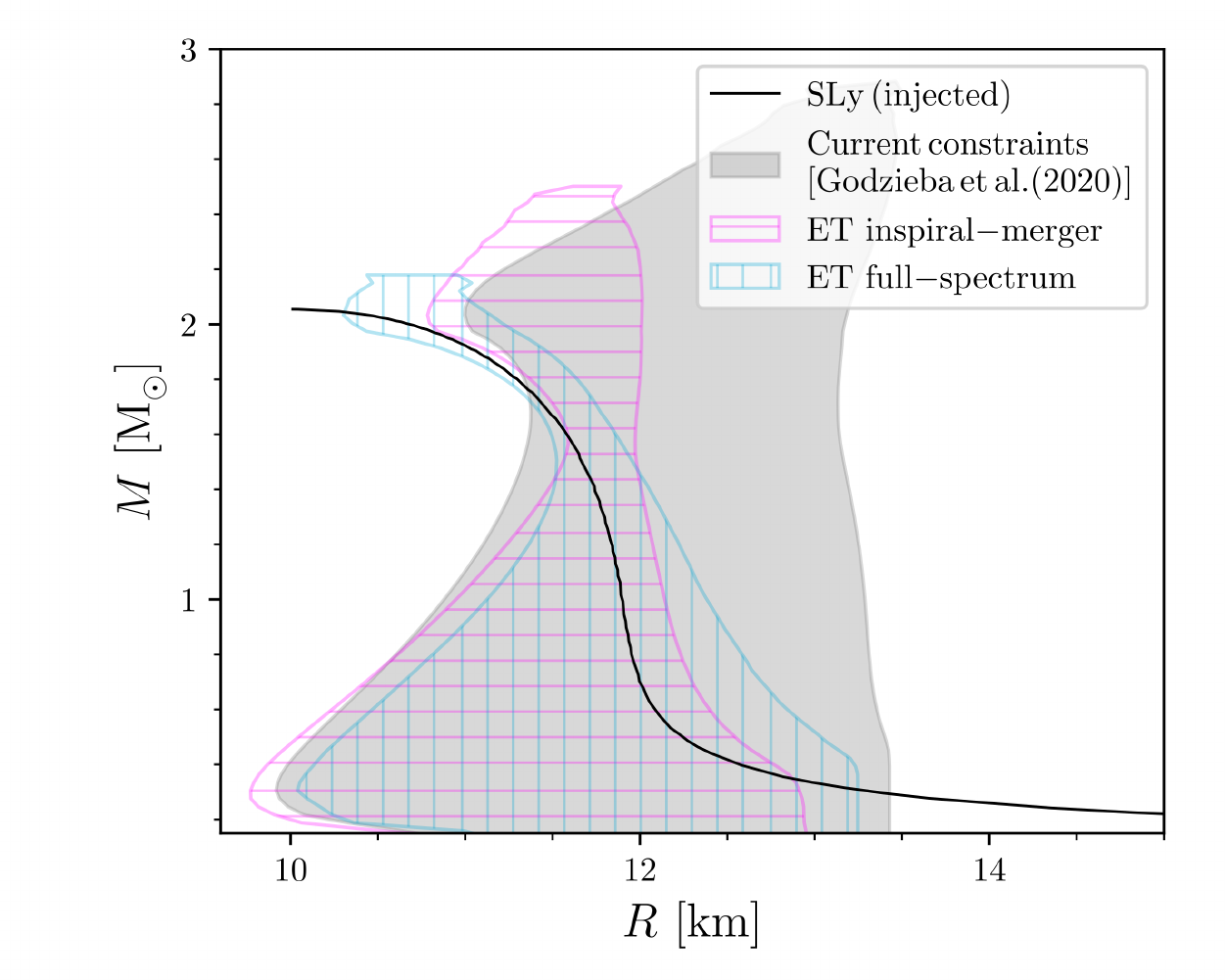}
	\caption{Mass-radius diagram constraints from a single 
	full-spectrum Einstein Telescope (ET) BNS observation
	with PM SNR 10 (total SNR $180$).
	The gray area (prior) corresponds to the two-million EOS
        sample of Ref.~\cite{Godzieba:2020tjn}.
        The magenta and cyan areas are the $90\%$ credibility regions given
        by inspiral-merger and inspiral-merger-PM inferences
        respectively. The full-spectrum (cyan) posterior agrees with the 
	injected EOS (black). }
	\label{fig:eos_samples}
\end{figure}

\paragraph{EOS constraints and mass-radius diagram.---}
To illustrate the potential impact of this approach, we show how a single
detection of PM signal at SNR 10 (total SNR 180 and 
luminosity distance of $120~{\rm Mpc}$)
can constrain the mass-radius relation for NSs.
The analysis makes use of the two-million EOS sample of
Ref.~\cite{Godzieba:2020tjn} that
are consistent with current constraints on the EOS of dense matter.
In particular, all EOSs predict maximum NS mass above the maximum NS
mass from pulsar observations, $\Mmax>1.97~\Msun$~\cite{Antoniadis:2013pzd}, and are consistent with the upper bound on the
tidal coupling constant from GW170817~\cite{Abbott:2018exr,Abbott:2018wiz,TheLIGOScientific:2017qsa}.
The only additional assumptions are the validity of general relativity
and that the EOS is causal up to the central density of the maximum mass NS.
We take the SLy binary as fiducial dataset and re-weight the EOS samples with the posteriors of the masses and 
the tidal parameters from the inspiral-merger and with the maximum
density and minimum radius posteriors from the PM measurement.

Figure~\ref{fig:eos_samples} shows the $90\%$ credibility regions
			  of the posterior distributions in the mass-radius diagram.
The inspiral-merger posteriors give the strongest constraint. 
They are mostly informative at the EOS at densities $\rho \simeq
2\rhosat$ corresponding to the individual NS components of the binary
\cite{Agathos:2015uaa}.
This inference leads to a measurement of the
maximum NS mass of $\Mmax=2.13^{+0.20}_{-0.14}\,\Msun$
($90\%$ credibility level), 
consistent with the injected value ($2.05\,\Msun$).
However, the EOS posterior shows a biased behavior 
for high mass values, i.e. $M \gtrsim 1.5\,\Msun$,
excluding the SLy sequence from the 90\% credibility region.
This shows that the inspiral-merger signal does not 
directly constrain the high-density EOS
and the inferred $\Mmax$ represents an extrapolation
based on the EOS representation.
The inclusion of PM information strengthens the agreement 
with the injected EOS at higher densities
allowing a measurements of $\Mmax=2.04^{+0.08}_{-0.06}\,\Msun$,
which agrees with the injected value 
and carries an error lower than $7\%$.
The improved constraint 
reduces of ${\sim}60\%$ the
$p(\rho)$ posterior area of the initial EOS sample leading, 
in particular, to tight pressure constraint at fiducial densities:
$\log_{10}\left[p(2 \rhosat) / ({\rm dyn}\,{\rm cm}^{-2}) \right] = 34.52^{+0.04}_{-0.04}$ 
and 
$\log_{10}\left[p(4 \rhosat) / ({\rm dyn}\,{\rm cm}^{-2}) \right] = 35.39^{+0.04}_{-0.03}$ 
 ($90\%$ credibility level).
For a detection at the sensitivity threshold (PM SNR 7.5) we find that 
the measured maximum NS mass is $\Mmax=2.07^{+0.15}_{-0.09}\,\Msun$ with a 
relative error of roughly ${\sim}12\%$.


\paragraph{Outlook.---} 
This work shows that next-generation GW observatories can deliver
strong constraints on the extreme-density EOS from a single,
full-spectrum detection of a BNS. In particular, the detection of the
PM signal from the merger remnant significantly enhances the measurement of the NS maximum
density and mass. 
The sensitivity threshold for PM BNS
transients is around a luminosity distance ${\sim}150\,{\rm Mpc}$, 
consistently with the recent estimates of Ref.~\cite{Abbott:2020uma}. 
According to our results, the multiple observation of about five PM BNS
transients at sensitivity threshold can lead to a measurement of the 
maximum NS mass with an error of ${\sim}5\%$~\cite{Torres-Rivas:2018svp,Easter:2020ifj}.
This information
  provides narrow observational constraints that would significantly
  inform nuclear models in the very-high-density regimes,
  \ie~$\rho > 3\rhosat$,
  which are unreachable conditions for modern nuclear experiments.
  Remarkably, the inclusion of the 
  PM inference contributes in reducing
  observational errors and methodological biases.

In real observations, EOS constraints at the accuracy level reported
here can be obtained by incorporating the PM signal in EOS inferences like those performed in
\cite{Landry:2018prl,Abbott:2018wiz,LIGOScientific:2019eut,Essick:2019ldf} for inspiral-merger signals.
In this context, the development of accurate EOS-insensitive relations
and the determination of their validity (or their breaking, \eg~\cite{Bauswein:2018bma,Raithel:2022orm}) 
via high-precision NR
simulations is a key step to improve the reliability of the GW measurement. 
Going beyond the Fisher matrix approach employed here, future work
will consider analyses employing Bayesian methods also for the
inspiral-merger.
Full spectrum analyses are currently possible given the 
availability of complete inspiral-merger-PM waveform models
\cite{Breschi:2019srl}. 
The latter include correlations on the mass and tidal coupling
constant between inspiral-merger and PM, that will further improve the
estimates of this work. 
However, the most urgent issue for the high-SNR
inspiral-mergers that will be observed by third-generation detectors
remains waveform systematics in the inference of tidal properties \cite{Gamba:2020wgg,Samajdar:2018dcx}. 
Nevertheless, the computational challenges related to full-spectrum Bayesian analyses might be confronted with parallel methods \cite{Smith:2019ucc,Breschi:2021wzr} and acceleration
techniques for the likelihood evaluation, \eg~ \cite{Smith:2016qas,Zackay:2018qdy,Schmidt:2020yuu,Cornish:2021lje}.


\begin{acknowledgments}
  \paragraph{Acknowledgments.---}
  The authors thank Francesco Zappa, Aviral Prakash and Andrea Camilletti 
  for providing unreleased NR data
  and Ssohrab Borhanian for 
  providing additional documentation for {\scshape gwbench}.
  Moreover, 
  we acknowledge important discussions in the 
  LIGO-Virgo {\it Extreme matter} group and within the Virgo-EGO collaboration, 
  in particular we thank
  Nikolaos Stergioulas, Katerina Chatziioannou and Jocelyn Read.
  MB and SB acknowledge 
  support by the EU H2020 under ERC Starting
  Grant, no.~BinGraSp-714626.
  MB acknowledges support from the Deutsche Forschungsgemeinschaft
  (DFG) under Grant No. 406116891 within the Research Training Group
  RTG 2522/1. 
  DR acknowledges funding from the U.S. Department of Energy, Office of
  Science, Division of Nuclear Physics under Award Number(s)
  DE-SC0021177 and from the National Science Foundation under Grants No.
  PHY-2011725, PHY-2020275, PHY-2116686, and AST-2108467.
  Virgo is funded by the French Centre National de Recherche Scientifique (CNRS), 
  the Italian Istituto Nazionale della Fisica Nucleare (INFN)
  and the Dutch Nikhef, with contributions by Polish and
  Hungarian institutes.
  Simulations were performed on {\scshape ARA},
  a resource of Friedrich-Schiller-Universt\"at Jena supported in part
  by DFG grants INST 275/334-1 FUGG, INST 275/363-1 FUGG and EU H2020
  BinGraSp-714626. 
  The {\bajes} software together with the {\nrpm} model 
  are publicly available at:
  \url{https://github.com/matteobreschi/bajes}.
  The NR simulations employed in this work are
  publicly available at:
  \url{http://www.computational-relativity.org/}.
\end{acknowledgments}


%

\end{document}